\documentclass{aastex7}

\begin{document}

\title{Reconnection Jets Detected during the Footpoint Drift of a CME Flux Rope: \\New Signatures of Ongoing 3D Arcade–Flux Rope Reconnection}

\author{Renjie Liu}
\affiliation{School of Physics and Astronomy, Yunnan University, Kunming, 650500,  People’s Republic of China}
\affiliation{Department of Astronomy, Key Laboratory of Astroparticle Physics of Yunnan Province, Yunnan University, \\Kunming 650091, People’s Republic of China}
\affiliation{Shanghai Astronomical Observatory, Chinese Academy of Sciences, 80 Nandan Road, Shanghai 200030, People’s Republic of China}
\email{}

\author[0000-0001-7866-4358]{Hechao Chen}
\affiliation{School of Physics and Astronomy, Yunnan University, Kunming, 650500,  People’s Republic of China}
\affiliation{Department of Astronomy, Key Laboratory of Astroparticle Physics of Yunnan Province, Yunnan University, \\Kunming 650091, People’s Republic of China}
\email{hechao.chen@ynu.edu.cn}  
\correspondingauthor {hechao.chen@ynu.edu.cn}

\author[0000-0002-7153-4304]{Chun Xia}
\affiliation{School of Physics and Astronomy, Yunnan University, Kunming, 650500,  People’s Republic of China}
\affiliation{Department of Astronomy, Key Laboratory of Astroparticle Physics of Yunnan Province, Yunnan University, \\Kunming 650091, People’s Republic of China}
\email{}  



\begin{abstract}
Utilizing multi-wavelength data from the Solar Dynamics Observatory, we report the first imaging detection of reconnection jets driven by the footpoint drift of an erupting hot-channel flux rope (HFR) during an X1.6-class solar flare. Our results demonstrate that these jets are inherent byproducts of the HFR eruption, exhibiting strong spatiotemporal coupling with the HFR footpoint drift and the morphological evolution of the associated flare ribbon hooks. Unlike standard coronal jets, these ejections lack an inverted-Y morphology and show no association with underlying photospheric flux cancellation or emergence. They manifest as clusters of small-scale, collimated ejecta (180–200 km s$^{-1}$) featuring co-existing cool and hot emission components visible in both cool (304 and 171~\AA) and hot (94~\AA) EUV channels. The imaging observations reveal that these jets are triggered by successive 3D arcade--flux rope (\textit{ar--rf}) reconnection as the expanding HFR interacts with the ambient coronal side arcades. The morphological and dynamic complexity of these observed jets implies that the ongoing \textit{ar--rf} reconnection in solar eruptions is episodic and bursty in nature. Our estimates show that a decrease of $\sim$ 9G in localized magnetic field strength is required to power these individual jets. As direct signatures of 3D magnetic reconnection, these observed jets pinpoint the exact site and characterize the nature of the process, providing new observational insight into 3D numerical simulation extensions of the standard flare model. Further observational case studies of these jets are required to fully understand their prevalence across solar eruptions.

\end{abstract}

\keywords{Solar activity(1475) ---Solar magnetic fields (1503) --- Solar magnetic reconnection(1504) --- Solar flares(1496) ---Solar coronal mass ejections (310)}


\section{INTRODUCTION} 

 {Solar flares and coronal mass ejections (CMEs) are the most energetic magnetic explosions in the solar atmosphere, which sometimes lead to strong disturbances in the near-Earth space environment and pose a hazard to satellite communications and space missions. These large-scale solar eruptions have been intensively studied over the past decades, and they have been well depicted by a standard flare model, i.e., the “CSHKP” model \citep[see the review of][and its references]{2003NewAR..47...53L}.  
In this model, a magnetic flux rope (MFR) is considered the core structure driving solar eruptions. The eruption begins when the MFR loses its equilibrium and rises, triggering magnetic reconnection within the vertical current sheet in its wake; this process releases pre-stored magnetic energy and reduces overlying magnetic constraints \citep[see more details in][]{2003NewAR..47...53L}. As a result, the magnetically unleashed MFR erupts rapidly to blow out a CME, while the simultaneous energy dissipation downward along reconnected field lines produces a solar flare. \citet{2012A&A...543A.110A,2013A&A...549A..66A} and \citet{2013A&A...555A..77J} successfully extended the CSHKP model into three dimensions (3D) using magnetohydrodynamical (MHD) simulations with the line-tying and zero-$\beta$ approximations. In the 3D extension of the standard flare model, \citet{2019A&A...621A..72A} first described and predicted a new reconnection geometry called arcade-flux rope (\textit{ar--rf}) reconnection converting arcade and rope to rope and flare loop, which has never been expected/reported in the traditional 2D model. This \textit{ar--rf} reconnection process happens between a CME flux rope leg and overlying/adjacent  coronal inclined arcades, which is predicted to erode the CME flux rope on one side while enlarging it on the other, leading to unexpected footpoint drifts of the CME flux rope during the eruption.  We hereafter refer to this footpoint drift  process of CME flux rope via \textit{ar--rf} reconnection as ``\textit{Aulanier footpoint drift effect}''.

Following the prediction of  \textit{Aulanier footpoint drift effect}, a series of observational studies have reported supportive evidence for its occurrence during real solar eruptions within the same year \citep{2019ApJ...887...71D,2019ApJ...887..118C,2019ApJ...885...83L,2019ApJ...883...96Z}. For instance, \citet{2019A&A...621A..72A} first reported the expanding-then-contracting evolution of flare hooks in two eruptive flares as described in \textit{ar--rf}  reconnection process. \citet{2019ApJ...887...71D} reported a long-distance shift of one leg of a filament and a growing ribbon hook that first swept over and finally enclosed the footpoints of pre-eruptive coronal arcades during a filament eruption. \citet{2019ApJ...887..118C} reported the simultaneous drift of both footpoints of an erupting flux rope near two conjugated coronal dimmings during an M-class flare; these drifting footpoints were anchored to slipping flare kernels that moved along the ribbon extremities, forming two nearly closed ribbon hooks. They underscored that this footpoint drift simultaneously caused the magnetic flux within the western dimming region to drop to approximately 38\% of its initial value, suggesting that \textit{ar--rf}  reconnection may play a critical role in the magnetic flux evolution of CME/ICME flux ropes. Moreover, \citet{2019ApJ...883...96Z} reported that during an M-class eruptive flare, a flare ribbon hook underwent rapid formation, disappearance, and subsequent reappearance in a different location, indicating a large shift between the footpoints of the pre-eruptive and erupting flux ropes. In the slow eruption process of a giant quiescent filament, \citet{2019ApJ...885...83L} also  observed a clear transformation of the legs of the filament's bright strands into hot flare loops above the sweeping flare ribbon hooks via \textit{ar--rf}  reconnection. These previous findings jointly indicate that the \textit{ar--rf}  reconnection geometry  can produce more complex observational phenomena related to the footpoint evolution of erupting CME flux ropes than what is expected from the line-tied assumption of magnetic field lines in traditional 2D models. Later on, the \textit{Aulanier footpoint drift effect} and its potential impacts have been gradually considered as new constraints that need to be taken into account in several aspects of subsequent observational studies \citep[see more details in the recent review of][]{2025SoPh..300..139D}, i.e., investigations of flux rope footpoint evolution \citep[e.g.,][]{2020ApJ...895...18B,2022ApJ...930..130L}, the morphology and origin of flare arcades \citep{2021ApJ...909L...4L}, high-speed slipping motions of flare kernels \citep{2025NatAs...9...45L}, falling filament mass flows along flux rope's legs \citep{2024ApJ...974..205S,2025ApJ...988...14O}, and outflows inside conjugate coronal dimmings \citep[e.g.,][]{2021ApJ...906...62L}, as well as attempts at the magnetic flux estimation and footpoint identification of CME flux ropes \citep[e.g.,][]{2020Innov...100059X,2021SSPMA..51i9531C,2023NatAs...7..815G}. 
Increasing observational evidence supporting the \textit{Aulanier footpoint drift effect} is updated in Solar Nugget hosted on HelioWiki\footnote{\url{https://heliowiki.smce.nasa.gov/wiki/index.php/The_Aulanier_Effect:_drifting_footpoints_of_CME_flux_ropes}}.

It is worth noting that the zero-$\beta$ CSHKP 3D simulation extension of \citet{2019A&A...621A..72A} can not describe the plasma dynamics and particle acceleration possibly caused by the  \textit{ar--rf}  reconnection. In real solar eruptions, if  \textit{ar--rf}  reconnection occurs among coronal inclined side arcades and CME flux rope's legs, observational signatures of its associated particle acceleration and plasma ejecta should, in theory, be detectable in the low corona. However, such observational clues have remained elusive and are rarely reported. Recently, \citet{2024A&A...690A.241Z} investigated the origin of groups of slowly positively drifting radio bursts (caused by accelerated electrons shot towards locations of higher plasma density) during a C8.7 flare whose flare kernels exhibited obvious slipping motions at the same time. Their results suggest that these radio bursts were produced by large-scale magnetic reconnection between an inflating sigmoid and a neighboring dome-like structure,  but it is unclear whether these signatures are closely related to the occurrence of \textit{ar--rf} reconnection.  More recently, \citet{2025A&A...698A.301J} reported the first spectroscopic signatures of blueshifts and line broadening at the reconnection interface between an erupting filament and overlying coronal arcades. Their observed reconnection process leads to the dissipation of coronal loops and the formation of a newborn hot flux rope, supporting the \textit{ar–rf} reconnection scenario described in \textit{Aulanier footpoint drift effect}.

In this paper, we report a type of reconnection jet triggered in the impulsive phase of the X1.6 flare (SOL2014-09-10T17:45) in solar active region (AR) NOAA 12158, utilizing multi-wavelength imaging data from the Solar Dynamics Observatory. These reconnection jets are found to be closely associated with the eruption and subsequent footpoint drift of an inflating coronal hot-channel flux rope. We investigate the formation process of these reconnection jets and their potential physical connection to the \textit{Aulanier footpoint drift effect} described in the 3D extended CSHKP model. Their distinctions from typical coronal jets and nanojets are also discussed.
The layout of this paper is as follows: Section 2 describes the instruments and data. In Section 3, we present our main observational results and interpretations. Finally, Section 4 discusses these reconnection jets and section 5 summarizes our results.

\section{Instruments and data} \label{sec:data}
 {The Atmospheric Imaging Assembly \citep[AIA;][]{2012SoPh..275...17L} on board the  {Solar Dynamics Observatory \citep[SDO;][]{2012SoPh..275....3P}} provides full-disk images of the Sun in 10  {EUV/UV} filters  {in a wide temperature range of 0.06 to 20 MK}. The temporal cadence and the spatial resolution of AIA EUV(UV)  {images} are 12 s(24 s) and  {1.$^{\prime\prime}$5, respectively}. AIA EUV images at 131 Å, 94 Å, 304 Å, and 171 Å, together with UV 1600 Å images, were employed in this study.  
The Helioseismic and Magnetic Imager \citep[HMI;][]{2014SoPh..289.3483H} on board SDO measures the full-disk photospheric magnetic fields at 6173 Å, providing routine line-of-sight (LOS) and vector magnetograms.  All these imaging data taken at different times are aligned to an appropriate reference time to remove  the solar rotation. 


\section{Results} \label{sec:observations}

The X1.6 flare occurred in AR NOAA 12158 and was accompanied by a spectacular high-speed CME  on 2014 September 10. The progenitor of this event, a hot-channel flux rope (HFR), was well-captured within the AR core by SDO. Its formation, evolution, and eruption have attracted extensive attention in numerous previous studies, particularly concerning the hot channel's high-temperature origin and plasma composition \citep{2016ApJ...823L...4C,2020ApJ...900L..18F}, its associated flare-induced chromospheric evaporation \citep{2015ApJ...811..139T}, flare ribbon dynamics \citep{2016ApJ...823...41D,2015ApJ...804L...8L,2016ApJ...823...62Z,2019A&A...621A..72A,2023NatAs...7..815G} and quasi-periodic pulsations \citep{2015ApJ...807...72L}, as well as its related precursor activities \citep{2016ApJ...823L..19Z,2017A&A...598A...3Z,2022RAA....22a5019S}. In the present study, we focus on the reconnection jets from the leg of the erupting HFR (see Figure \ref{fig:1}(a)), which has never been reported or investigated. 

\subsection{The footpoint drift of the erupting flux rope due to  \textit{ar--rf}  reconnection} \label{subsec:eruption}

\begin{figure*}[ht!]
\centering
\includegraphics[width=0.7\linewidth]{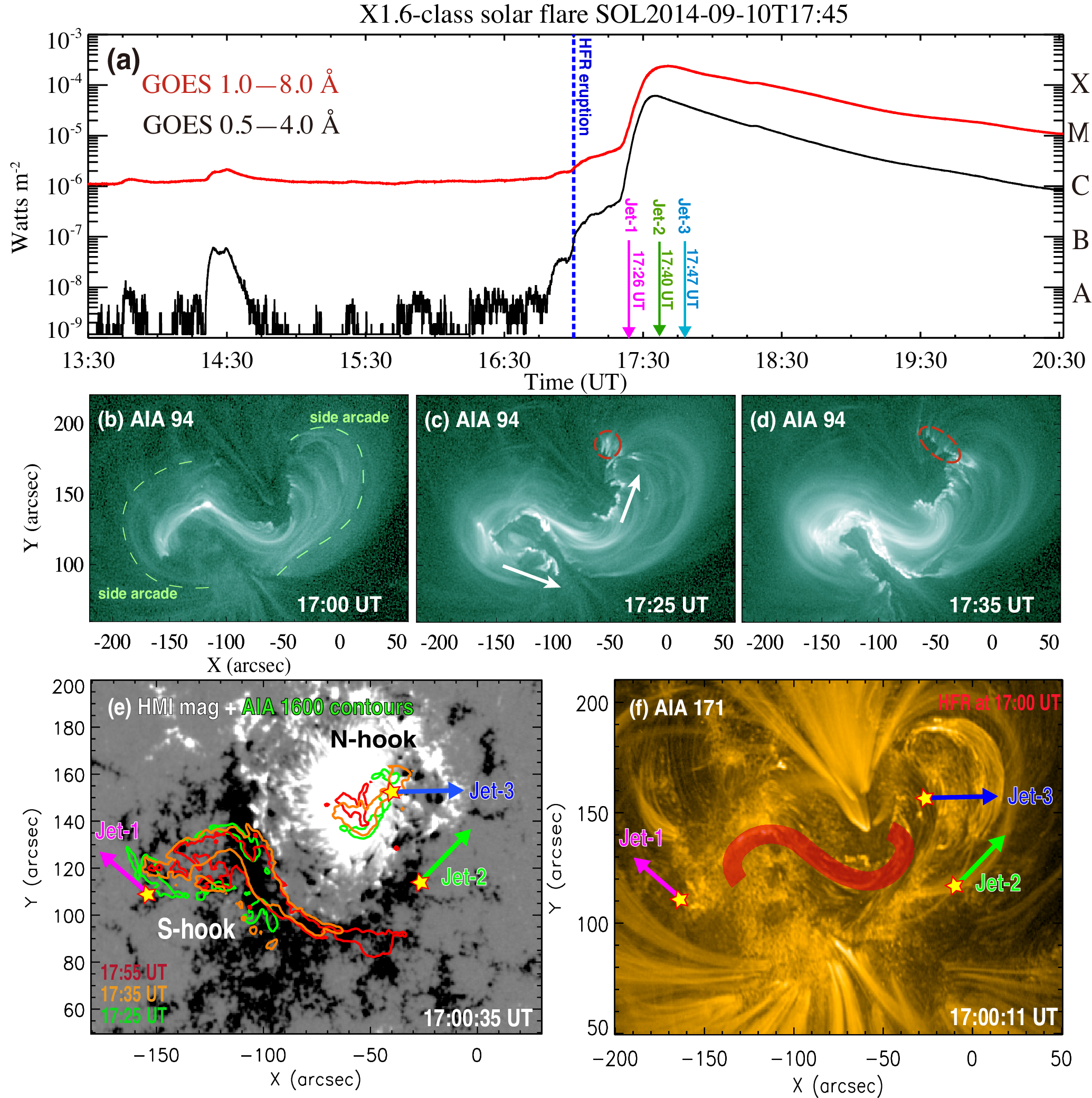}
\caption{Overview of the X1.6-class solar flare (SOL2014-09-10T17:45).
(a) GOES-15 soft X-ray flux in the 0.5–4 Å (black) and 1–8 Å (red) channels. The vertical dashed line and colored arrows denote the onset of the hot-channel flux rope (HFR) eruption (blue) and the initiation of various reconnection jets, respectively. 
(b)–(d) Eruption of the HFR. Arrows in panel (c) indicate the slipping directions of the footpoints, while the red dashed circles highlight new magnetic footpoints during the eruption. (e) HMI line-of-sight magnetogram overlaid with contours of the flare ribbons during the HFR eruption, as recorded in AIA 1600 Å images at different times.
(f) Spatial distribution of three groups of reconnection jets detected within the ambient coronal arcades outside the HFR. The red sigmoidal structure outlines the pre-eruption HFR, derived from the AIA 94 Å image at 17:00 UT. In panels (e) and (f), colored arrows and star symbols denote the initiation sites of reconnection jets (Jet-1 to Jet-3) and their ejection directions, respectively. }
\label{fig:1}
\end{figure*}

At 17:00 UT, the HFR initially appeared as an inverted-S sigmoidal HFR structure (Figure \ref{fig:1}(b)), very likely formed through pre-flare reconnection between two J-shaped loop systems as reported by \citep[see observational examples in][]{2018ApJ...869...78C,2019ApJ...887..118C}. Two bundles of pre-existing coronal side arcades (green dashed lines in Figure \ref{fig:1}(b)) envelop the outer boundary of the sigmoidal HFR, which was later found to be involved the triggering of reconnection jets in Section 3.2. Following its rapid expansion at $\sim$17:20 UT, the HFR erupted, triggering an X1.6 flare that peaked at $\sim$17:45 UT (Figure \ref{fig:1}(a)). A striking feature of this eruption was the prominent slipping signatures at the HFR footpoints from 17:20 UT to 17:50 UT: the two HFR legs underwent rapid migration—toward the southwest and northeast, respectively (as marked by the white arrows in Figure \ref{fig:1}(c))—forming two distinct flare-ribbon hooks (Figure \ref{fig:1}(e)). These hooks, termed the S-hook and N-hook, enclosed a pair of conjugate core dimming regions and were anchored in negative and positive polarities, respectively (Figure \ref{fig:1}(e)). 

The footpoint drift near the S-hook and its associated flare ribbon dynamics have been extensively studied \citep{2016ApJ...823...41D, 2019A&A...621A..72A, 2023NatAs...7..815G,2015ApJ...804L...8L}. In particular, \citet{2019A&A...621A..72A} confirmed that \textit{ar--rf} reconnection between the inflating HFR and adjacent coronal side arcades drives the unique ``expansion-and-shrink" motion of the S-hook, a key prediction of  \textit{Aulanier footpoint drift effect}. 
Similarly, signatures of footpoint drift and flare ribbon slipping motions were also observed near the N-hook (as shown by the colored flare ribbon contours in Figure \ref{fig:1}(e)). These features are consistent with the predictions of  \textit{Aulanier footpoint drift effect}, supporting the occurrence of \textit{ar--rf} reconnection near the N-hook during the eruption.

\begin{figure*}[ht!]
\centering
\includegraphics[width=0.8\linewidth]{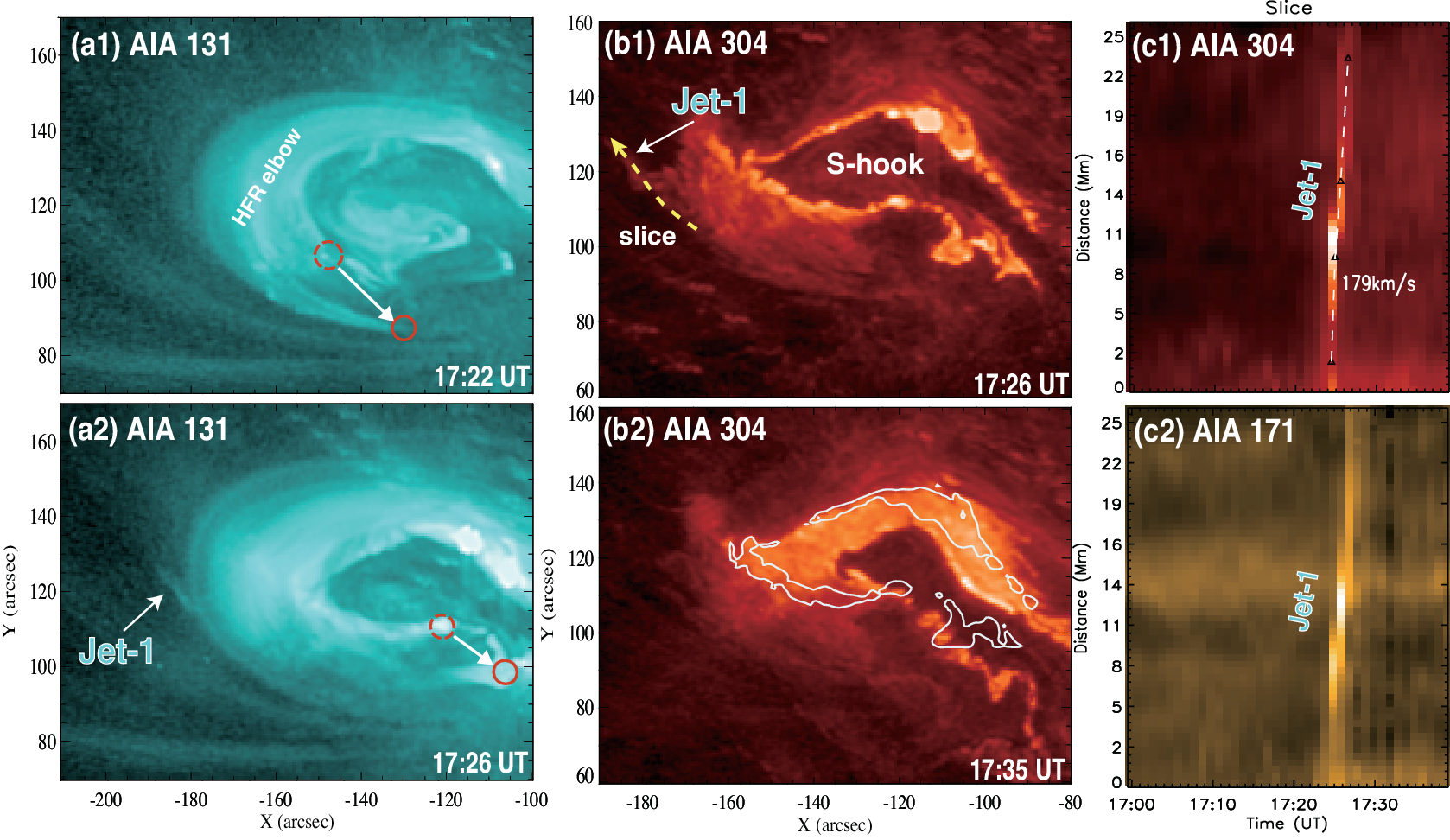}
\caption{Reconnection jet Jet-1 observed near the S-hook region. (a1–a2) and (b1–b2) EUV images showing the expansion of the HFR's eastern elbow and its associated S-hook, respectively. The red dashed and solid circles highlight the footpoint of a flux rope bundle prior to and after its drifting. The white contours of the S-hook at Jet-1 onset (from panel b1) are overlaid on panel (b2) at the time of jet termination. (c1–c2) Time-distance diagrams of Jet-1, sliced along the yellow arrow indicated in panel (b1). A 2-second animation using AIA 304 and 171 Å passbands shows reconnection Jet-1 from 17:00:44–17:46:12 UT.
\\(One animation of this figure is available.)
}
\label{fig:2}
\end{figure*}

\subsection{Reconnection Jets detected during the HFR footpoint drift \label{subsec:jet}}

During the HFR eruption, three distinct reconnection jet groups were identified above the N-hook and S-hook,  as well as along right-side coronal arcades near the S-hook’s tip, respectively (as indicated by colored arrows in  Figures \ref{fig:1}(e) and (f)). We termed them as Jet-1 to Jet-3, according to their occurrence moments  (see Figure \ref{fig:1}(a)), all of which were spatio-temporally coupled with the HFR footpoint drift  in the flare impulsive/peak phase. In the following sections, we investigate the morphological evolution and kinematic properties of these jets to elucidate their connection to the \textit{ar--rf} reconnection geometries described in \textit{Aulanier footpoint drift effect}.

\subsubsection{Reconnection jets above S-hook} \label{subsec:ar Jet}
Jet-1 was detected at the  HFR's eastern elbow expanding above the S-hook during the onset of the HFR expansion stage (17:24--17:28 UT).
At this stage, the nascent HFR grew and expanded, likely driven by pre-flare reconnection beneath it. This preflare reconnection induced a slow footpoint drift of the HFR (marked by red circles and white arrows in Figures \ref{fig:2}(a1)--(a2)), leading to the formation and subsequent expansion of the nascent S-hook (Figures \ref{fig:2}(b1)--(b2)). Consequently, the inflating HFR elbow may collide with the ambient coronal side arcades (left-side green dashed line in Figure \ref{fig:1}(b)), triggering \textit{ar--rf} reconnection at their interface. As a result, Jet-1 was rapidly ejected from the HFR elbow region (Figures \ref{fig:2}(a2) and (b1)). This transient jet is visible across the AIA 131~\AA, 304~\AA, and 171~\AA\ passbands (see the accompanying movie). Time-distance analysis in Figures \ref{fig:2}(c1)--(c2) reveals that Jet-1 propagated to a distance of 18--20~Mm, with a projected velocity of $\sim$180~km~s$^{-1}$ and a lifetime of $\sim$3~min.

\subsubsection{Reconnection jets detected along right-side arcades}\label{subsec:ar Jet}

Jet-2 erupted along the right-side coronal arcades at 17:39:11 UT, nearly concurrent with the flare peak at $\sim$17:40 UT. 
At first glance, Jet-2 appeared to emanate from the intersection of two surrounding coronal side arcades (``AB'' and ``CD"  in Figure \ref{fig:5}(a1)) via their direct reconnection. This direct reconnection between arcades "AB'' and "CD'' is expected to produce simultaneous brightenings at all four footpoints of the two interacting arcades. 
However, a detailed examination instead indicates that Jet-2 was associated with multiple sequential EUV brightenings at distinct footpoints  of several coronal side arcades,  with significant time lags between these brightenings (see  “A'',  “B'',  “C'', and  “D'' in Figures \ref{fig:5}(a3–a9) and the following description).  Therefore, we suggest that the eruption of Jet-2 is most likely triggered by successive \textit{ar--rf} reconnection events.

Figure \ref{fig:5} and accompanying animation demonstrate the detailed onset process of Jet-2, its associated EUV brightenings, and the subsequent formation of flare loops.
Prior to the onset of Jet-2, two sets of preexisting coronal loops labeled “AB” and “CD” are visible at 17:00 UT. Footpoints “A” and “C” lie within the N-hook region, while footpoints “B” and “D” anchor the vicinity of the S-hook. Weak compact brightenings first arise near locations “A” and “C” at 17:15 UT as the HFR becomes activated and begins to erupt. At 17:25 UT, the northern footpoint of the HFR undergoes abrupt brightening to form the N-hook, which subsequently drifts gradually westward and interacts with the ambient side arcades. Following this westward drift, successive localized EUV enhancements appear at sites “A”, “C”, and “D” over the interval 17:30–17:39 UT (see Figure \ref{fig:5}(a4-a7)). At 17:39:11 UT, Jet-2 initiates abruptly and propagates outward along arcade “AB”, marked by the green arrow in Figure \ref{fig:5}(a4). The launch of Jet-2 coincides with a concentrated EUV brightening at the original footpoint of the preexisting arcade “AB”, circled in red and labeled “1” in Figure \ref{fig:5}(a4). Meanwhile, the coronal arcade hosting the jet rapidly shifted its footpoint from the initial position (marked by the red circle in Figure \ref{fig:5}(a4))  to a new southern footpoint area (marked by the red circle in Figure \ref{fig:5}(a5)). This rapid footpoint migration is clearly visible in the supplementary movie.

\begin{figure*}[ht!]
\centering
\includegraphics[width=0.8\linewidth]{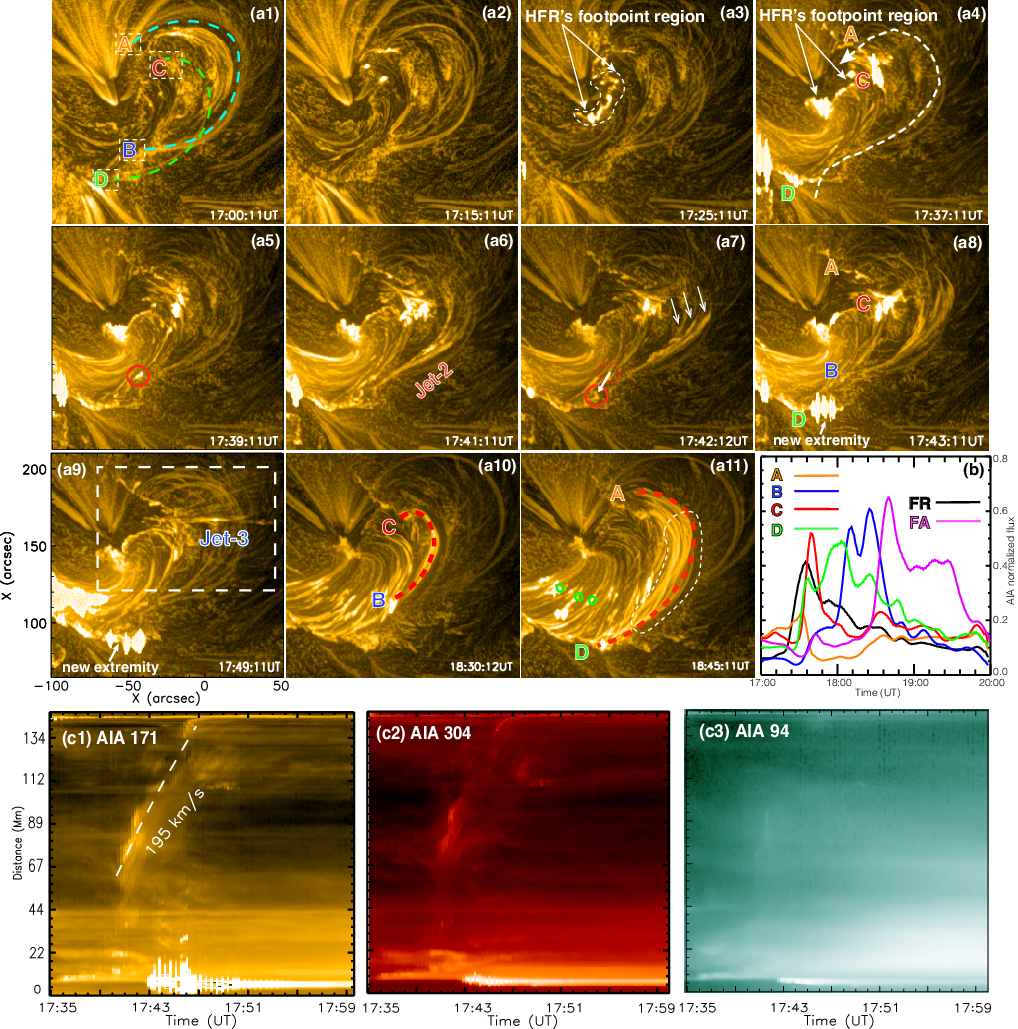}
\caption{(a1–a11) Onset of Jet-2 and the associated topology change of the coronal side arcades (colored dashed lines, including pre-jet side arcades: ``AB" and ``CD", as well as post-jet newborn flare arcades: ``CB" and ``AD" ) observed in SDO/AIA 171 Å images enhanced by the MGN code. The paired white arrows in (a3) and (a4) mark the expanding northern footpoint brightening region of the HFR. Red solid/dashed circles in (a5) and (a7) represent the footpoint features of the coronal side arcades before and after drift, which participated in the launch of Jet-2. White arrows in (a7) indicate the embedded sub-jets within Jet-2. The white dashed rectangle in panel (a7) outlines the field of view (FOV) containing Jet-3, which is displayed in Figure \ref{fig:4}. (b) Temporal evolution of the AIA 171 Å light curves integrated over the arcade footpoint regions ( “A”, “B”, “C”, and “D”, marked by white boxes in (a1)),the HFR footpoint region (marked by white outline in (a3)), and the newborn side flare arcade region (marked by white outline in (a11)), respectively. (c1–c3) Time-distance diagrams across multiple EUV channels, sliced along the Jet-2 trajectory (indicated by the white dashed arrow in a4).  A 6-second animation utilizing AIA 171, 304, and 94 Å passbands displays reconnection Jet-2 and Jet-3 over 17:00:11–18:50:12 UT.
\\(An animation of this figure is available.)}
\label{fig:5}
\end{figure*}

After its launch, the Jet-2 becomes more conspicuous by 17:41:11 UT and continues northwestward, reaching the northern terminus of the right-side coronal arcades by 17:43:41 UT. 
Notably, Jet-2 consists of numerous fine-scale sub-jets, clearly resolved in Figures \ref{fig:5}(a5)–(a7) and the supplementary animation. Each sub-jet features an extended jet spire propagating along the flanking arcades. During its full evolution, successive sub-jets are launched along the coronal side arcades, combining to form the large-scale structure of Jet-2. Time–distance diagrams presented in  Figures \ref{fig:5}(c1)–(c3) show that Jet-2 is most prominent in the relatively cool 304, 171\ \AA\ passbands, but is barely detectable in the hotter 94\ \AA\ channel.  The projected velocity of Jet-2 is estimated to be approximately 200 km s$^{-1}$. 

As shown in Figure \ref{fig:5} (a8), while Jet-2 is still ongoing, a new EUV brightening soon appears near the location “D” at 17:43:11 UT, forming a new flare extremity that stretches outward from the S-hook. Concurrently, Jet-3 erupts from the EUV brightening site “C” and becomes most prominent at 17:49:11 UT. The detailed initiation of Jet-3 will be discussed in the following section. The AIA 171 \AA\ flux curves in Figure \ref{fig:5} (b), integrated over the arcade footpoint regions (“A”, “B”, “C”, and “D”), the HFR footpoint region, and the newborn side flare arcade region, indicate that multiple reconnection episodes likely participated in the initiation of both Jet-2 and Jet-3. Specifically, the EUV enhancement at site “C” first appears at $\sim$17:39 UT and decays by 17:48 UT, whereas the brightening at site “D” exhibits a temporal lag of approximately three minutes. This time delay between the two signatures suggests they arise from multiple reconnection episodes, not a single isolated reconnection event between arcades ``AB'' and ``CD''.

Approximately half an hour after the eruptions of Jet-2 and Jet-3, two bundles of newly formed side flare arcades “CB” and “AD” gradually appeared within the jet region at 18:30 UT and 18:45 UT, respectively (also see the decaying peak feature in the purple AIA light curve in Figure \ref{fig:5}(b)). Analogous to the atypical flare arcades reported by \citet{2019ApJ...887..118C} and  \citet{2021ApJ...909L...4L}, these side flare arcades are most probably generated by the \textit{ar--rf} reconnection geometry during the Aulanier footpoint drift effect. Consequently, their morphology differs drastically from canonical post-flare loops (marked with green circles in Figure \ref{fig:5}(a8)) predicted by the two-dimensional standard eruption model \citep{2003NewAR..47...53L}.


\subsubsection{Reconnection jets above N-hook}\label{subsec:ar Jet}

Figure \ref{fig:4} illustrates the detailed triggering process of Jet-3. Jet-3 erupts above the N-hook during 17:43–17:52 UT, revealing a direct physical connection to one of the conjugate core coronal dimmings produced in the wake of the erupting HFR (evident from the outward-expanding purple contours in Figures \ref{fig:4}(b1)–(b3)).  The onset of Jet-3 is preceded by a compact EUV coronal enhancement at 17:41 UT, highlighted by the yellow star in Figure \ref{fig:4}(a3). This brightening forms at the magnetic interface where the rising HFR impinges upon a cluster of preexisting flanking coronal arcades (denoted by the right green dashed line in Figure \ref{fig:1}(b) and (e)),  providing a favorable site for \textit{ar-rf} reconnection described in the  \textit{Aulanier footpoint drift effect}. This interface coincides with the edge of the northern core dimming zone (Figures \ref{fig:4}(a2)–(a3)), outlined by the N-hook flare ribbon and generated by substantial plasma evacuation along the flux rope leg during its rapid inflation. As the core dimming expands significantly westward, Jet-3 is swiftly launched from this magnetic interface and propagates toward the west. 

\begin{figure*}[ht!]
\centering
\includegraphics[width=0.8\linewidth]{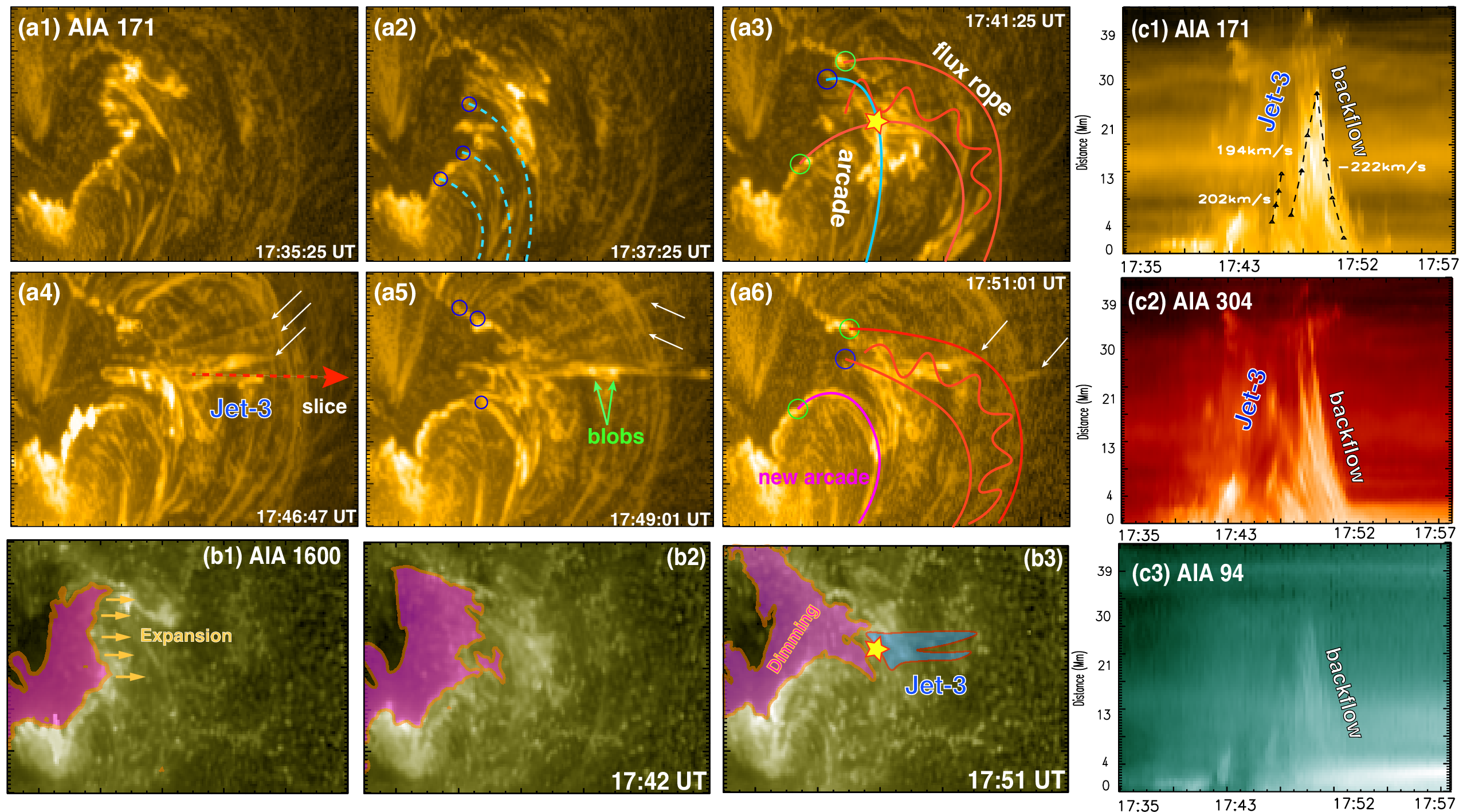}
\caption{(a1–a6) Initiation of Jet-3 and associated coronal dimming expansion near the N-hook region (SDO/AIA 171 Å images enhanced by MGN code). Blue and green circles in (a2, a3, a5, a6) mark the footpoints of the coronal side arcades and the HFR before and after reconfiguration, respectively. Panels (a3) and (a6) illustrate the proposed magnetic topology reconfiguration between the erupting HFR leg (red) and ambient coronal side arcades (cyan), yielding a newly formed side arcade (purple). White arrows in (a4–a6) indicate spicule-like sub-jets nested within Jet-3; green twin arrows in (a5) mark moving blobs. (b1–b3) AIA 1600 Å images showing N-hook evolution, overlaid with contours of the expanding coronal dimming (purple) and Jet-3 region (blue) derived from the 171 Å channel. (c1–c3) Multi-wavelength time-distance diagrams of Jet-3, sliced along the red dashed path in (a4). The FOV of (a1-a6) and (b1-b3) is denoted by the white rectangle in Figure \ref{fig:5}(a9).  A 3-second animation using AIA 171, 304, 1600, and 94 Å passbands shows reconnection Jet-3 from 17:00:11–18:01:55 UT. 
\\(An animation of this figure is available.) }
\label{fig:4}
\end{figure*}

As marked by white arrows in Figures \ref{fig:4}(a4)–(a6), Jet-3 manifests as a cluster of collimated, spicular plasma ejecta containing multiple sub-jets embedded within its primary spire. Jet-3 attains a maximum projected length of $\sim$35~Mm before prominent plasma backflow develops along its ejection path. Time-distance diagrams in Figure \ref{fig:4}(c1)--(c3) reveal an ejection velocity of 194--202~km~s$^{-1}$ and a subsequent backflow velocity of $\sim$ -222~km~s$^{-1}$, suggesting that the plasma remained trapped on newly reconnected field lines. In addition, Jet-3 is detectable over a broad temperature regime across the 304, 171, and 94~\AA\ EUV passbands. After the jet erupts, the ambient magnetic topology undergoes a rapid dynamic rearrangement via footpoint interchange between the eruptive HFR and the preexisting right-side coronal arcades (Figures \ref{fig:4}(a3) and (a6)). The original footpoints of both magnetic structures disappear entirely and are substituted by new footpoints shifted northward from the initial interaction site, consistent with a magnetic reconnection pattern described as “as one falls, another rises.”

\subsection{Possible Interpretation} \label{sec:NLFFF}

According to previous imaging observations,  a possible scenario is proposed in Figure \ref{fig:8} to explain the triggering of these reconnection jets, within the framework of the 3D extension of the standard flare model \citep{2019A&A...621A..72A}. In this scenario, the HFR (yellow contours) expands and collides with low-lying flanking coronal arcades along its eastern and western sides. This collision drives a series of complex \textit{ar–rf} reconnections between the opposite-polarity magnetic field components of the expanding HFR (yellow lines) and the preexisting lateral coronal arcades (blue/cyan lines). Due to these \textit{ar–rf}  reconnection events, the legs of the HFR fully “escape” the lateral magnetic confinement imposed by the original ambient arcades (blue lines), enabling the flux rope to proceed into its final eruptive stage while its footpoints migrate to new positions together with slipping hook-shaped flare ribbons.

We propose that Jet-1 originates from an isolated \textit{ar–rf} reconnection event above the S-hook, where the eastern elbow of the outward-drifting HFR interacts with the left-side coronal arcades. By contrast, Jet-2 is plausibly launched through multiple \textit{ar–rf} reconnection episodes, since its onset is correlated with sequential EUV brightening signatures at discrete locations. Resolving such fast and sequential \textit{ar–rf} reconnection processes remains a challenge for current instrumental time cadences; even so, one of these reconnection episodes clearly links Jet-2’s initiation to the right-side arcades, concurrent with footpoint interchange across these lateral magnetic structures.
Notably, Jet-2 and Jet-3 are not independent phenomena. The \textit{ar–rf} reconnection underlying Jet-2 drives the HFR footpoints to drift northward. This northward migration creates favorable magnetic conditions for further \textit{ar–rf} reconnection episodes above the N-hook, which subsequently drives Jet-3. Due to multiple \textit{ar–rf} reconnection episodes, multiple sets of newly formed side flare arcades (orange/red lines) and their corresponding footpoint EUV enhancements (red contours) develop at the respective sites. We caution that, during the initiation of Jet-2, this scenario primarily accounts for the \textit{ar–rf} reconnection geometry; nevertheless, the involvement of \textit{arcade–arcade} reconnection cannot be entirely ruled out.

\begin{figure*}[ht!]
\centering
\includegraphics[width=1.0\linewidth]{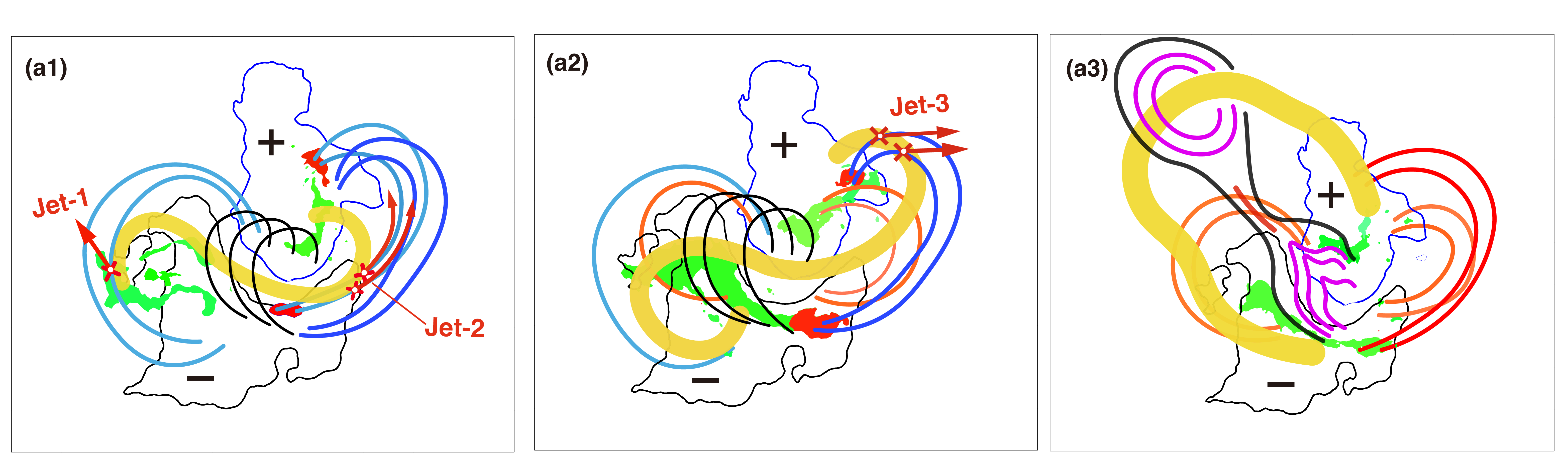}
\caption{Cartoons illustrate the onset of reconnection jets during different phases of the HFR eruption. Magnetic polarities (black and blue contours) and the associated flare ribbons (green-filled contours) are extracted from HMI line-of-sight (LOS) magnetograms and AIA 1600 Å observations, respectively. In the cartoon, the HFR is represented by the yellow channel, while standard post-flare loops are denoted by purple curves. Pre-jet and post-jet coronal side arcades are indicated by cyan/blue and orange/red curves, respectively. }
\label{fig:8}
\end{figure*}

\section{Discussion}

\subsection{Reconnection jets: new and direct signatures for ongoing \textit{ar--rf} reconnection} \label{sec:NLFFF}

The \textit{ar--rf} reconnection geometry was first described and predicted in 3D simulation extensions of the standard flare model. Occurring between the leg of an erupting flux rope and the overlying or ambient coronal side arcades, this process dynamically reconfigures the morphology and magnetic flux of the erupting CME flux rope, significantly influencing solar flare evolution \citep[see the review of][]{2025SoPh..300..139D}. To date, several observational signatures have been identified as indirect evidence of \textit{ar--rf} reconnection: (1) the rapid expansion-then-contraction and/or slipping motions of flare ribbon hooks during eruption \citep{2019A&A...621A..72A,2019ApJ...887..118C,2019ApJ...883...96Z,2025NatAs...9...45L}; (2) the continuous drifting of filament/flux rope footpoints accompanied by clear coronal topological changes \citep{2019ApJ...887...71D,2019ApJ...887..118C,2019ApJ...885...83L}; and (3) the formation of saddle-shaped flare arcades above the ribbon hooks \citep{2021ApJ...909L...4L}. While these signatures provide compelling support, they primarily represent consequential remnants that are insufficient to capture the ongoing, dynamic nature of the \textit{ar--rf} reconnection process itself. 

Unlike prior studies, our current work reports imaging detection of \textit{ar--rf} reconnection jets for the first time during the footpoint drift of an erupting CME flux rope. These observed jets provide direct and immediate markers of ongoing  \textit{ar--rf} reconnection, allowing us to not only pinpoint the reconnection sites but also characterize the physical evolution of the reconnection process itself. As described in Section 3, these \textit{ar--rf} reconnection jets were characterized by a series of small-scale energy releasing events rather than a single, monolithic eruption. Morphologically, these jets exhibit significant complexity, comprising clusters of collimated plasma ejecta (see Figures \ref{fig:2} and  \ref{fig:4}). In particular,  Jet-3 exhibits a more intermittent and episodic nature, manifesting as a series of spicular jets with distinct blob-like features embedded within their spires (see Figure \ref{fig:4} (a5)). This indicates that the \textit{ar--rf}  reconnection between the opposite-polarity magnetic field components of the HFR leg and inclined coronal side arcades is episodic and bursty in nature. 
Moreover, the chronological sequence of Jet-1 through Jet-3 is closely dictated by the specific magnetic configuration. The \textit{ar–rf} reconnection events appear to first occur either between the HFR and the lower-lying arcades closer to the active region core or between the lower arcades themselves, before subsequently taking place between the higher-altitude arcades and the HFR legs.
 
Similarly, \citet{2025A&A...698A.301J} also recently detected prominent spectroscopic blueshifts and line broadening at the reconnection interface between an erupting filament and its overlying coronal arcades. Their observed blueshifts are approximately 200 km s$^{-1}$, consistent with the speed of the \textit{ar–rf} reconnection jets identified in our work. Their observed reconnection process yields coronal loop dissipation and the formation of a new hot flux rope, but it tends to suppress the initial filament eruption, with no accompanying obvious CME. By contrast, the reconnection jets we report arise as the legs of the CME flux rope successfully break through the lateral magnetic confinement by surrounding side coronal arcades through \textit{ar–rf} reconnection. Notably, the imaging detection of \textit{ar–rf} reconnection jets presented in this work has never been reported in prior studies of eruptive CME flux rope events. This discrepancy stands in contrast to the substantial footpoint drift observed in our current event, which—according to 3D simulation extensions of the standard flare model—suggests that \textit{ar--rf} reconnection should occur over a much broader region and an extended duration. This discrepancy in observational detections may indicate that specific physical conditions are required either for the initiation of these reconnection jets or for their observational detection in real solar eruptions. We suggest that three primary factors may favor the occurrence and detection of \textit{ar--rf} reconnection jets: (1) the pre-existence of abundant coronal side arcades enveloping the eruptive core structure, providing the necessary magnetic configuration to facilitate widespread \textit{ar--rf} reconnection; (2) a relatively low-lying \textit{ar--rf} reconnection site, allowing the resulting jets to be more easily detected in EUV passbands; and (3) very high-cadence imaging observations, which are essential to capture these transient and weak features.
In the future, more observational investigations and full MHD simulations are warranted to determine whether \textit{ar--rf} reconnection jets are a ubiquitous feature of solar eruptions or a unique phenomenon requiring specific topological conditions.

\subsection{Reconnection jets: comparisons with coronal jets and nanojets} \label{sec:comparison}

Coronal jets are small-scale, collimated plasma ejections triggered by magnetic reconnection between open and closed (or twisted) magnetic fields. They typically exhibit an inverted-Y morphology comprising a collimated spire (averaging $\sim$8~Mm wide and $\sim$50~Mm long) and a broader base containing an asymmetric, localized bright point \citep[for reviews, see][]{2016SSRv..201....1R,2021RSPSA.47700217S}. Extensive observations reveal that these events typically originate at mixed-polarity regions featuring clear photospheric flux emergence or cancellation \citep[e.g.,][]{2007A&A...469..331J,2011ApJ...738L..20H,2012ApJ...745..164S,2015ApJ...815...71C,2016ApJ...821..100S,2016ApJ...832L...7P,2017ApJ...851...67S,2019ApJ...877...61M,2023ApJ...942...86Y,2024ApJ...962L..38D,2025ApJ...987..193Y}.
Besides, nanojets represent a distinct class of smaller, short-lived, and high-velocity ejecta recently discovered via high-resolution observations \citep[e.g.,][]{2020ApJ...899...19C,2021NatAs...5...54A,2025ApJ...995...94C,2025ApJ...985L..12G,2025A&A...702A.189T}. Typically measuring only $\sim$0.5 Mm in width and $\sim$1.0 Mm in length with lifetimes under 15 s and typical velocities spanning 50–800 km s$^{-1}$, nanojets are believed to be driven by reconnection between opposite-polarity components under strong guide fields, propagating roughly perpendicular to the ambient field \citep[also see simulations:][]{2021NatAs...5...54A,2025A&A...699A.106S,2026arXiv260503055C}.

Unlike traditional coronal jets, the reconnection jets reported here are inherent to the 3D coronal reconnection process during solar eruptions rather than being isolated events. Spatiotemporally coupled with the HFR footpoint drift during the flare impulsive and peak phases, these eruption-induced jets manifest morphologically as clusters of bursty, spicular ejecta. Crucially, they lack the distinct jet bases and characteristic inverted-Y shape typically associated with standard coronal jets. Furthermore, these ejections are initiated in the upper atmosphere near the N-hook and S-hook regions (indicated by colored arrows in Figures \ref{fig:1}(e) and (f)), where no obvious photospheric flux emergence or cancellation exists below. Taken together, these observational characteristics demonstrate that these jets are triggered in the transition-region or low coronal environment via \textit{ar--rf}  reconnection, distinguishing them from traditional coronal jets.

The reconnection jets reported here exhibit typical projected velocities of 180–200 km s$^{-1}$, spire lengths of 5–35 Mm, and lifetimes of 3–8 min. Notably, these physical parameters fall within the characteristic range of standard coronal jets, despite their distinct underlying triggering mechanisms. Moreover, like standard coronal jets (but unlike recently reported nanojets), these ejections are field-aligned features that propagate along either the ambient side arcades or the expanding HFR leg. Assuming these propagation speeds are comparable to the local Alfvén speed, we estimate the localized magnetic field strengths required to power each individual jet to be approximately 9 G.
In addition, these ejections have multi-thermal emission signatures, characterized by the co-existence of cool and hot plasma components, closely resembling the emission nature of both standard coronal jets \citep{2007A&A...469..331J,2017ApJ...851...67S} and nanojets \citep{2026arXiv260503055C}. Such a multi-temperature nature implies that the \textit{ar--rf} reconnection occurs in the lower corona, where the hot HFR interacts and reconnects with the cooler surrounding side arcades.

\section{Summary} \label{sec:summary}

In this work, we report the first detection of eruption-induced reconnection jets driven by the footpoint drift of an eruptive CME flux rope during an X1.6-class solar flare. Spatiotemporally coupled with the expansion and contraction of the conjugate flare ribbon hooks and variations in the coronal core dimmings, these jets serve as immediate, dynamic indicators of ongoing 3D \textit{ar–rf} reconnection. We investigated their initiation, dynamics, and associated coronal configuration changes using multi-wavelength imaging observations. Unlike traditional coronal jets driven by photospheric flux cancellation or emergence, these ejections are inherent byproducts of the erupting CME flux rope interacting with ambient coronal side arcades. Morphologically, they lack an inverted-Y structure, instead manifesting as episodic clusters of discrete, multi-thermal spicular or blob-like features. While these jets share similar kinematic parameters with standard coronal jets, their fine-scale, intermittent nature reveals that \textit{ar–rf} reconnection in the low corona is fundamentally bursty and episodic, providing new observational insight into 3D extensions of the standard CSHKP flare model. More case studies are needed to confirm that these reconnection jets constitute a common behavior of the 3D standard flare model.

\begin{acknowledgments}

This work is supported by the NSFC grant (12573061) and the Xingdian Talents Support Program of Yunnan Province (Young Talent Program: XDYC-QNRC-2023-0255), as well as the Yunnan Provincial Basic Research Project (202401CF070165). 
The authors would like to thank the referee for very constructive suggestions that improved this manuscript, and Prof. Yuandeng Shen for his helpful comments. L. R. J. completed this research as his undergraduate thesis under the supervision of C. H. C. at Yunnan University. 
SDO is a mission for NASA’s Living With a Star program. \end{acknowledgments}

\bibliography{sample7}{}
\bibliographystyle{aasjournalv7}



\end{document}